\documentclass[aps,prd,preprint,tightenlines,nofootinbib]{revtex4}
\usepackage{graphicx}
\usepackage{dcolumn}
\usepackage{bm}

\newcommand{\etal}{{\it et al.}}

\begin{document}

\preprint{\tighten\vbox{\hbox{\hfil EFI 13-19}}}
\title
{\LARGE Leptonic Decays of Charged Pseudoscalar Mesons $-$ 2013}

\author{Jonathan L. Rosner}
\affiliation{Enrico Fermi Institute, University of Chicago, Chicago, IL 60637}
\author{and Sheldon Stone}
\affiliation{Department of Physics, Syracuse University, Syracuse, NY 13244 }

\bigskip

\date{\today}

\begin{abstract}
We review the physics of purely leptonic decays of $\pi^\pm$, $K^\pm$,
$D^{\pm}$, $D_s^\pm$, and $B^\pm$ pseudoscalar mesons.  The measured decay
rates are related to the product of the relevant weak-interaction-based CKM
matrix element of the constituent quarks and a strong interaction parameter
related to the overlap of the quark and antiquark wave-functions in the meson,
called the decay constant $f_P$. The interplay between theory and experiment is
different for each particle. Theoretical predictions of $f_B$ that are needed
in the $B$ sector can be tested by measuring $f_{D^+}$ and $f_{D_s^+}$ in the
charm sector.  The lighter $\pi^{\pm}$ and $K^{\pm}$ mesons provide stringent
comparisons between experiment and theory due to the accuracy of both the
measurements and the theoretical predictions. An abridged version of this
review was prepared for the Particle Data Group's 2014 edition \cite{Previous}.
\end{abstract}
\maketitle

\section{Introduction}
\begin{figure}[hbt]
\centering
\includegraphics[width=3.5in]{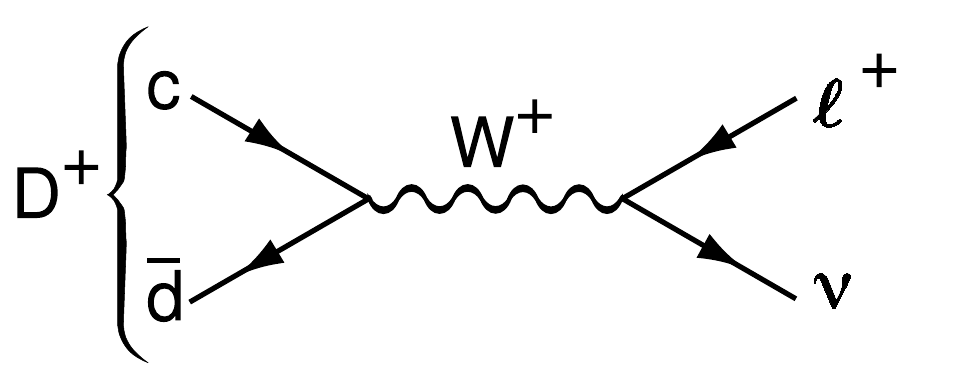}\vskip -0.02mm
\caption{The annihilation process for pure $D^+$ leptonic decays in the
Standard Model.
 } \label{Ptoellnu}
\end{figure}

Charged mesons formed from a quark and antiquark can decay to a
lepton-neutrino pair when these objects annihilate via a virtual
$W$ boson. Fig.~\ref{Ptoellnu} illustrates this process for the
purely leptonic decay of a $D^+$ meson.

Similar quark-antiquark annihilations via a virtual $W^+$ to
the $\ell^+ {\nu}$ final states occur for the $\pi^+$, $K^+$, $D_s^+$, and
$B^+$ mesons.  (Charge-conjugate particles and decays are implied.) Let $P$ be
any of these pseudoscalar mesons.  To lowest order, the decay width is
\begin{equation}
\Gamma(P\to \ell\nu) = {{G_F^2}\over 8\pi}f_{P}^2\ m_{\ell}^2M_{P}
\left(1-{m_{\ell}^2\over M_{P}^2}\right)^2 \left|V_{q_1
q_2}\right|^2~. \label{equ_rate}
\end{equation}

\noindent Here $M_{P}$ is the $P$ mass, $m_{\ell}$ is the $\ell$
mass, $V_{q_1 q_2}$ is the Cabibbo-Kobayashi-Maskawa (CKM) matrix
element between the constituent quarks $q_1 \bar q_2$ in $P$, and
$G_F$ is the Fermi coupling constant. The parameter $f_P$ is the
decay constant, proportional to the matrix element of the axial current
between the one-$P$-meson state and the vacuum, and is related to the
wave-function overlap of the quark and antiquark.

The decay $P^\pm$ starts with a spin-0 meson, and ends up with a
left-handed neutrino or right-handed antineutrino.  By angular
momentum conservation, the $\ell^\pm$ must then also be left-handed
or right-handed, respectively. In the $m_\ell = 0$ limit, the decay
is forbidden, and can only occur as a result of the finite $\ell$
mass.  This helicity suppression is the origin of the $m_\ell^2$
dependence of the decay width. Radiative corrections are needed
when the final charged particle is an electron or muon \cite{Bradcor}.

Measurements of purely leptonic decay branching fractions and
lifetimes allow an experimental determination of the product
$\left|V_{q_1 q_2}\right| f_{P}$. If the CKM element is well known
from other measurements, then $f_P$ can be well measured. If, on the
other hand, the CKM element is not well measured, having
theoretical input on $f_P$ can allow a determination of the CKM
element.   The importance of measuring $\Gamma(P\to \ell\nu)$
depends on the particle being considered. In the case of the $B^-$ the 
measurement of $\Gamma(B^-\to\tau^-\overline{\nu})$ provides an indirect 
determination of $|V_{ub}|$ provided that $f_B$ is provided by theory.
In addition, $f_B$ is crucial for using measurements of $B^0$-$\overline{B}^0$
mixing to extract information on the fundamental CKM parameters.  Knowledge
of $f_{B_s}$ is also needed, but it cannot be directly
measured as the $B_s$ is neutral, so the violation of the SU(3)
relation $f_{B_s} = f_B$ must be estimated theoretically. This
difficulty does not occur for $D$ mesons as both the $D^+$ and
$D_s^+$ are charged, allowing the measurement of SU(3)
breaking and a direct comparison with theory.

For $B^-$ and $D_s^+$ decays, the existence of a charged Higgs boson
(or any other charged object beyond the Standard Model, SM) would modify
the decay rates; however, this would not necessarily be true for
the $D^+$ \cite{Hou,Akeroyd}.  More generally, the ratio of $\tau \nu$
to $\mu \nu$ decays can serve as one probe of lepton universality
\cite{Hou,Hewett}.

\overfullrule 0pt As $|V_{ud}|$ has been quite accurately measured
in superallowed $\beta$ decays \cite{Vud}, with a value of
0.97425(22) \cite{BM}, measurements of $\Gamma(\pi^+ \to \mu^+{\nu})$ yield a
value for $f_{\pi}$. Similarly, $|V_{us}|$ has been well measured in
semileptonic kaon decays, so a value for $f_{K}$ from $\Gamma(K^-
\to \mu^- \bar{\nu})$ can be compared to theoretical calculations.
Lattice gauge theory calculations, however,  have been
claimed to be very accurate in determining $f_K$, and these have
been used to predict $|V_{us}|$ \cite{Jutt}.

\section{Charmed mesons}
Our review of current measurements starts with the charm system.  Measurements
have been made for $D^+\to\mu^+\nu$, and  $D_s^+\to\mu^+\nu$ and $D_s^+\to\mu^+
\nu$ and $\tau^+\nu$. Only an upper limit has been determined for $D^+\to\tau^+
\nu$. Both CLEO-c and BES have made measurements of  $D^+$ decay using $e^+e^-$
collisions at the $\psi(3770)$ resonant energy where $D^-D^+$ pairs are
copiously produced. They fully reconstruct one of the $D$'s, say the $D^-$.
Counting the number of these events provides the normalization for the
branching fraction measurement. They then find a candidate $\mu^+$, and then
form the missing-mass squared, $MM^2=\left(E_{\rm CM}-E_{D^-}\right)^2-
\left(\overrightarrow{p}_{\!\rm CM}-\overrightarrow{p}_{\!D^-}-
\overrightarrow{p}_{\!\mu^+}\right)^2$, taking into account their knowledge of
the center-of-mass energy, $E_{\rm CM}$,  and momentum, $p_{\rm CM}$, that
equals zero in $e^+e^-$ collisions.  A peak at zero $MM^2$ infers the
existence of a missing neutrino and hence the $\mu^+\nu$ decay of the $D^+$.
CLEO-c does not explicitly identify the muon, so their data consists of a
combination of $\mu^+\nu$ and $\tau^+\nu$, $\tau^+\to\pi^+\nu$ events. This
permits them to do two fits; in one they fit for the individual components, and
in the other they fix the ratio of $\tau^+\nu/\mu^+\nu$ events to be that given
by the SM expectation. Thus, the latter measurement should be used for SM
comparisons and the other for new physics searches.  Our average uses the fixed
ratio value.  The measurements are shown in Table~\ref{tab:fDp}.

\begin{table}[htb]
\caption{Experimental results for ${\cal{B}}(D^+\to \mu^+\nu)$, ${\cal{B}}
(D^+\to \tau^+\nu)$, and $f_{D^+}$. Numbers for $f_{D_s^+}$ have been
extracted using updated values for masses and $|V_{cd}|$ (see text). Radiative
corrections are included. Systematic uncertainties arising from the $D^+$
lifetime and mass are included.
\label{tab:fDp}}
\begin{center}
\begin{tabular}{llccc}
 \hline\hline
& Experiment & Mode & ${\cal{B}}$ & $f_{D^+}$ (MeV)\\ \hline
& CLEO-c \cite{fD}& $\mu^+\nu$& $(3.93\pm
0.35\pm 0.09)\times 10^{-4}$ & $209.1 \pm 9.3\pm 2.5$\\
& CLEO-c \cite{fD}& $\mu^+\nu$+$\tau^+\nu$ & $(3.82\pm
0.32\pm 0.09)\times 10^{-4}$ & $206.2 \pm 8.6\pm 2.6$\\ 
& BES \cite{BESfD}& $\mu^+\nu$& $(3.74\pm
0.21\pm 0.06)\times 10^{-4}$ & $204.0\pm 5.7\pm 2.0$\\
\hline
& Average & $\mu^+\nu$ & $(3.76\pm 0.18)\times 10^{-4}$ & $204.6\pm 5.0$ \\
\hline
& CLEO-c \cite{CLEO-c} & $\tau^+\nu$ & $<1.2\times 10^{-3}$ &
\\ \hline\hline
\end{tabular}
\end{center}
\end{table}

To extract the value of $f_{D^+}$ we use the well-measured $D^+$ lifetime of 
1.040(7) ps. The value of  $|V_{cd}|$ is taken to equal to the value of
$|V_{us}|$ of 0.2252(9) \cite{BM}  minus higher order correction terms
\cite{Charles}, which results in $|V_{cd}|=0.2251(9)$.  The $\mu^+\nu$ results 
include a 1\% correction (lowering) of the rate due to the presence of the
radiative $\mu^+\nu\gamma$ final state based on the estimate by
Dobrescu and Kronfeld \cite{Kron}.

Before we compare this result with theoretical predictions, we discuss the
$D_s^+$. Measurements of $f_{D_s^+}$ have been made by several groups and are
listed in Table~\ref{tab:fDs} \cite{CLEO-c,Belle-munu,CLEO-rho,CLEO-CSP,%
Sanchez}.  We exclude older values obtained by normalizing to $D_s^+$ decay
modes that are not well defined. Many measurements, for example, used the
$\phi\pi^+$ mode. This decay is a subset of the $D_s^+\to K^+ K^- \pi^+$
channel which has interferences from other modes populating the $K^+K^-$ mass
region near the $\phi$, the most prominent of which is the $f_0(980)$. Thus
the extraction of the effective $\phi\pi^+$ rate is sensitive to the mass
resolution of the experiment and the cuts used to define the $\phi$ mass region
\cite{reason,Babar-munu}. 

To find decays in the $\mu^+ \nu$  signal channels, CLEO, BaBar and Belle rely
on fully reconstructing all the final state particles except for neutrinos and
using a missing-mass technique to infer the existence of the neutrino. CLEO
uses $e^+e^-\to D_sD_s^*$ collisions at 4170 MeV, while Babar and Belle use
$e^+e^-\to D Kn\pi D_s^*$ collisions at energies near the $\Upsilon(4S)$.
CLEO does a similar analysis as was done for the $D^+$ above.  Babar and Belle
do a similar $MM^2$ calculation by using the reconstructed hadrons, the photon
from the $D_s^{*+}$ decay and a detected $\mu^+$. To get the normalization they
do a $MM^2$ fit without the $\mu^+$ and use the signal at the $D_s^+$ mass
squared to determine the total $D_s^+$ yield. 

When selecting the $\tau^+\to\pi^+\bar{\nu}$ and $\tau^+\to\rho^+\bar{\nu}$
decay modes, CLEO uses both calculation of the missing-mass and the fact that
there should be no extra energy in the event beyond that deposited by the
measured tagged $D_s^-$ and the $\tau^+$ decay products. The $\tau^+\to
e^+\nu\bar{\nu}$ mode, however, uses only extra energy.  Babar and Belle also
use the extra energy to discriminate signal from background in their
$\tau^+\nu$ measurements.

\begin{table}[b]
\caption{Experimental results for ${\cal{B}}(D_s^+\to \mu^+\nu)$, ${\cal{B}}
(D_s^+\to \tau^+\nu)$, and $f_{D_s^+}$. Numbers for $f_{D_s^+}$ have been
extracted using updated values for masses and $|V_{cs}|$ (see text). Systematic
uncertainties for errors on the $D_s^+$ lifetime and mass are included;
radiative corrections have been included. Common systematic errors in each
experiment have been taken into account in the averages.
\label{tab:fDs}}
\begin{center}
\begin{tabular}{llccc}
 \hline\hline
& Experiment & Mode & ${\cal{B}}$ & $f_{D_s^+}$ (MeV)\\ \hline
& CLEO-c \cite{CLEO-c}& $\mu^+\nu$& $(5.65\pm
0.45\pm 0.17)\times 10^{-3}$ & $257.6\pm 10.3\pm 4.3$\\
& BaBar \cite{Sanchez}& $\mu^+\nu$& $(6.02\pm
0.38\pm 0.34)\times 10^{-3}$ & $265.9\pm 8.4\pm 7.7$\\
&Belle \cite{Belle-munu}
& $\mu^+\nu$ & $(5.31\pm 0.28\pm 0.20)\times 10^{-3}$ & $249.7
\pm 6.6 \pm 5.0$ \\
\hline
& Average & $\mu^+\nu$ & $(5.56\pm 0.24)\times 10^{-3}$ & $255.6\pm 5.9$ \\
\hline
& CLEO-c \cite{CLEO-c} & $\tau^+\nu~(\pi^+\overline{\nu})$ & $(6.42\pm 0.81\pm
0.18) \times 10^{-2}$ & $278.0\pm 17.5 \pm 4.4 $ \\
& CLEO-c \cite{CLEO-rho} & $\tau^+\nu~(\rho^+\overline{\nu})$ & $(5.52\pm 0.57
\pm 0.21)\times 10^{-2}$ & $257.8\pm 13.3 \pm 5.2 $ \\
& CLEO-c \cite{CLEO-CSP} & $\tau^+\nu~(e^+\nu\overline{\nu})$ &
$(5.30\pm 0.47\pm 0.22)\times 10^{-2}$ & $252.6\pm 11.2 \pm 5.6 $ \\
& BaBar \cite{Sanchez} & $\tau^+\nu~(e^+(\mu^+)\nu\overline{\nu})$ &
$(5.00\pm 0.35\pm 0.49)\times 10^{-2}$ & $245.4\pm 8.6 \pm 12.2 $ \\
&Belle \cite{Belle-munu}
& $\tau^+\nu~(\pi^+\overline{\nu})$  & $(6.04\pm 0.43^{+0.46}_{-0.40})
 \times 10^{-2}$ & $269.6\pm 9.6^{+10.4}_{-9.1} $ \\
&Belle \cite{Belle-munu}
& $\tau^+\nu~(e^+\nu\overline{\nu})$  & $(5.37\pm 0.33^{+0.35}_{-0.31})
 \times 10^{-2}$ & $254.2\pm 7.8^{+8.5}_{-7.6} $ \\
&Belle \cite{Belle-munu}
& $\tau^+\nu~(\mu^+\nu\overline{\nu})$  & $(5.86\pm 0.37^{+0.34}_{-0.59})
 \times 10^{-2}$ & $265.5\pm 8.4^{+7.9}_{-13.5}  $ \\ \hline
&Average & $\tau^+\nu$ & $(5.56\pm 0.22\times 10^{-2})$ & $258.3\pm 5.5$
\\ \hline\hline
\end{tabular}
\end{center}
\end{table}

We extract the decay constant from the measured branching ratios using the
$D_s^+$ mass of 1.96849(32) GeV,  the $\tau^+$ mass of 1.77682(16) GeV, and a
lifetime of 0.500(7) ps.  We use the first order correction $|V_{cs}| =
|V_{ud}| - |V_{cb}|^2/2$ \cite{Charles}; taking $|V_{ud}| = 0.97425(22)$
\cite{Vud}, and $|V_{cb}| =0.04$ from an average of exclusive and inclusive
semileptonic $B$  decay results as discussed in Ref.~\cite{Vcb}, and find
$|V_{cs}| = 0.97345(22)$.   CLEO has
included the radiative correction of 1\% in the $\mu^+\nu$ rate listed in the
Table \cite{Kron}~(the $\tau^+\nu$ rates need not be corrected). Other
theoretical calculations show that the $\gamma\mu^+\nu$ rate is a factor of
40--100 below the $\mu^+\nu$ rate for charm \cite{theories-rad}. As this is a
small effect we do not attempt to correct the other measurements.

The average decay constant cannot simply be obtained by averaging the values
in Table~\ref{tab:fDs} since there are correlated errors between the $\mu^+\nu$
and $\tau^+\nu$ values.  Table \ref{tab:both} gives the average values of
$f_{D_s}$ where the experiments have included the correlations.

\begin{table}[htb]
\caption{Experimental results for $f_{D_s^+}$ taking into account the common
systematic errors in the $\mu^+\nu$ and $\tau^+\nu$ measurements.}
\label{tab:both}
\begin{center}
\begin{tabular}{lc}
 \hline\hline
Experiment &$f_{D_s^+}$ (MeV)\\\hline
CLEO-c & $259.0\pm 6.2\pm 3.0$\\
BaBar & $258.4\pm 6.4\pm 7.5$\\
Belle & $257.8\pm 4.2\pm 4.8$\\\hline
Average of $\mu^+\nu+\tau^+\nu$ & $257.5\pm 4.6$
\\ \hline\hline
\end{tabular}
\end{center}
\end{table}

Our experimental average is
$$
f_{D_s^+}=(257.5\pm 4.6){\rm ~MeV}.
$$
Furthermore, the ratio of branching fractions is found to be 
\begin{equation}
R\equiv \frac{{\cal{B}}(D_s^+\to \tau^+\nu)}{{\cal{B}}(D_s^+\to \mu^+\nu)} =
10.0 \pm 0.6,
\end{equation}
where a value of 9.76 is predicted in the SM. Assuming lepton universality then
we can derive improved values for the leptonic decay branching fractions of
\begin{eqnarray}
{\cal{B}}(D_s^+\to \mu^+\nu)&=&(5.64\pm0.20)\times 10^{-3},~~{\rm and}
\nonumber\\ {\cal{B}}(D_s^+\to \tau^+\nu)&=&(5.51\pm0.20)\times 10^{-2}~.
\end{eqnarray}

\begin{table}[htb]
\caption{Theoretical predictions of $f_{D^+_s}$, $f_{D^+}$, and
$f_{D_s^+}/f_{D^+}$.  Quenched lattice calculations are omitted, while
PQL indicates a partially-quenched lattice calculation.
(Only selected results having errors are included.)
\label{tab:Models}}
\begin{center}
\begin{tabular}{llccc}
\hline\hline
& Model & $f_{D_s^+}$(MeV) & $f_{D^+}$(MeV) & $f_{D_s^+}/f_{D^+}$\\\hline
& Experiment (our averages)& $257.5 \pm 4.6$ & $204.6\pm5.0$& $1.258\pm 0.038$
 \\ \hline
  & Lattice (HPQCD) \cite{Lat:NaD} & $246.0\pm0.7\pm3.5$ & $208.3\pm1.0\pm3.3$
  & $1.187 \pm 0.004 \pm 0.012$ \\ 
 & Lattice (FNAL+MILC) \cite{Lat:Milc} &
$246.4\pm0.5\pm3.6$ & $209.2\pm3.0\pm 3.6$ & $1.175\pm0.019$ \\
&PQL \cite{Lat:Nf2}& $244\pm 8$&$197\pm 9$&
   $1.24\pm 0.03$\\
& QCD sum rules \cite{Bordes} & $205\pm 22$& $177\pm 21$& $1.16\pm 0.01\pm0.03$
\\
& QCD sum rules \cite{Lucha} & $245.3\pm15.7\pm4.5$ & $206.2\pm7.3\pm5.1$ &
 $1.193\pm0.025\pm0.007$ \\
 &QCD sum rules \cite{Narison} & $246\pm 6$ & $204\pm 6$  &$1.21\pm 0.04$ \\
 &QCD sum rules \cite{Wang} (I) & $241\pm12$ & $208\pm11$ & $1.16\pm0.07$ \\
 &QCD sum rules \cite{Wang} (II) & $258\pm13$ & $211\pm14$ & $1.22\pm0.08$ \\
 &QCD sum rules \cite{Gelhausen} & $238^{+13}_{-23}$ & $201^{+12}_{-13}$ &
  $1.15^{+0.04}_{-0.05}$ \\
& Field correlators \cite{Field} & $260\pm 10$& $210\pm 10$& $1.24\pm 0.03$\\
& Light front \cite{LF} & $268.3\pm 19.1$ & 206 (fixed) & $1.30\pm 0.04$\\
\hline\hline
\end{tabular}
\end{center}
\end{table}

The experimentally determined ratio of decay constants is $f_{D_s^+}/f_{D^+}
=1.258\pm 0.038$. Table~\ref{tab:Models} compares the experimental $f_{D_s^+}$
with theoretical calculations \cite{Lat:NaD,Lat:Milc,Lat:Nf2,Bordes,Lucha,
Narison,Field,LF}.  Most theories give values lower than the $f_{D_s^+}$
measurement. The discrepancy with the the models with the smallest quoted
uncertainties, both unquenched lattice calculations, are 2.0 standard
deviations with HPQCD \cite{Lat:NaD}, and 1.9 standard deviations with the
preliminary FNAL+MILC prediction \cite{Lat:Milc}.

Upper limits on $f_{D^+}$  and $f_{D_s}$ of 230 and 270 MeV, respectively, have
been determined using two-point correlation functions by Khodjamirian
\cite{Kho}.  The $D^+$ result is safely below this limit, while the average
$D_s^+$ result is also, but older results \cite{Previous} not used in our
average are often above the limit.

Akeroyd and Chen \cite{AkeroydC} pointed out that leptonic decay widths are
modified in two-Higgs-doublet models (2HDM).  Specifically, for the $D^+$ and
$D^+_s$, Eq.~(\ref{equ_rate}) is modified by a factor $r_q$ multiplying the
right-hand side \cite{AkeroydM}:

$$
r_q=\left[1+\left(1\over{m_c+m_q}\right)\left({M_{D_q}\over M_{H^+}}\right)^2
\left(m_c-\frac{m_q\tan^2\beta}{1+\epsilon_0\tan\beta}\right)\right]^2,
$$

\noindent where $m_{H^+}$ is the charged Higgs mass, $M_{D_q}$ is
the mass of the $D$ meson (containing the light quark $q$), $m_c$ is
the charm quark mass, $m_q$ is the light-quark mass, and $\tan\beta$
is the ratio of the vacuum expectation values of the two Higgs
doublets. In models where the fermion mass arises from coupling to more
than one vacuum expectation value $\epsilon_0$ can be non-zero, perhaps
as large as 0.01. For the $D^+$, $m_d \ll m_c$, and the change due to the
$H^+$ is very small. For the $D_s^+$, however, the effect can be substantial.

In order to investigate the possible presence of new physics we need to
specify a SM value of $f_{D_s^+}$.  We can only use a theory prediction. Our
most aggressive choice is that of the unquenched lattice calculation
\cite{Lat:NaD}, because it claims the smallest error.  Since the charged Higgs
would lower the rate compared to the SM, in principle, experiment gives a lower
limit on the charged Higgs mass.  However, the value for the predicted decay
constant using this model is 2.0 standard deviations {\it below} the
measurement.  If this small discrepancy is to be taken seriously, either (a)
the model of Ref.~\cite{Lat:NaD} is not representative; (b) no value of
$m_{H^+}$ in the two-Higgs doublet model will satisfy the constraint at 99\%
confidence level; or (c) there is new physics, different from the 2HDM, that
interferes constructively with the SM amplitude such as in the
R-parity-violating model of Akeroyd and Recksiegel \cite{Rviolating}.

To sum up, the standard model calculations are now consistent with the data and
new physics effects are small. Limits can be placed on new particles depending
on the specific model.

\section{\boldmath\bf The $B^-$ meson}
The Belle and BaBar collaborations have found evidence for $B^-\to\tau^-
\overline{\nu}$ decay in $e^+e^-\to B^-B^+$ collisions at the $\Upsilon(4S)$
energy.  The analysis relies on reconstructing a hadronic or semi-leptonic $B$
decay tag, finding a $\tau$ candidate in the remaining track and photon
candidates, and examining the extra energy in the event which should be close
to zero for a real $\tau^-$ decay to $e^- \nu \bar \nu$ or $\mu^- \nu \bar \nu$
opposite a $B^+$ tag. While the BaBar results have remained unchanged, Belle
did a re-analysis of their data using the hadronic $B$ decay sample. The
branching fraction changed from $1.79\,^{+0.56\,+0.46}_{-0.49\,-0.51}\times
10^{-4}$ \cite{BelleH} to $0.72^{+0.27}_{-0.25}\pm0.11\times10^{-4}$
\cite{BelleHnew}. This change demonstrates the difficulty of the analysis. It
is unfortunate that other results have not been updated.  The results are
listed in Table~\ref{tab:Btotaunu}.

\begin{table}[htb]
\caption{Experimental results for ${\cal{B}}(B^-\to \tau^-\overline{\nu})$.
\label{tab:Btotaunu}}
\begin{center}
\begin{tabular}{lllc} \hline\hline
&Experiment & Tag &${\cal{B}}$ (units of $10^{-4}$)\hfill\\
\hline
&Belle~\cite{BelleHnew}&Hadronic&$0.72^{+0.27}_{-0.25}\pm0.11$\\
&Belle~\cite{BelleS}&Semileptonic&$1.54\,^{+0.38\,+0.29}_{-0.37\,-0.31}$\\
&Belle~\cite{BelleHnew}&Average&$0.96 \pm 0.26$ \\\hline
&BaBar~\cite{BaBarH} & Hadronic & $1.83\,^{+0.53}_{-0.49}\pm0.24$\\
&BaBar~\cite{BaBarS} & Semileptonic & $1.7\pm 0.8\pm 0.2$\\
&BaBar~\cite{BaBarH}  & Average & $1.79 \pm 0.48$\\\hline
& &Our average & $1.14\pm0.23$\\
\hline\hline
\end{tabular}
\end{center}
\end{table}

There are large backgrounds under the signals in all cases. The systematic
errors are also quite large. Thus, the significances are
not that large.  Belle quotes 3.0$\sigma$ and 3.6$\sigma$ for their hadronic
and semileptonic tags, respectively, while BaBar quotes 3.3$\sigma$ and 2.3%
$\sigma$, again for hadronic and semileptonic tags. More accuracy would be
useful, especially to investigate the effects of new physics.

We extract a SM value using Eq.~(\ref{equ_rate}). Here theory provides a value
of $f_B=(190.6\pm 4.7)$ MeV \cite{fBl}.
We also need a value for $|V_{ub}|$. Here significant differences
arise between using inclusive charmless semileptonic decays and
the exclusive decay $B\to\pi\ell^+\nu$ \cite{ABS}.  The inclusive decays give
rise to a value of $|V_{ub}|=(4.41\pm 0.22)\times 10^{-3}$  while the exclusive
measurements yield $|V_{ub}|=(3.23 \pm 0.31)\times 10^{-3}$, where the errors
are dominantly theoretical \cite{KM}.  Their average,
enlarging the error in the standard manner because the results differ, is
$|V_{ub}|=(4.01\pm0.56)\times 10^{-3}$. Using these values and the PDG values
for the $B^+$ mass and lifetime, we arrive at the SM prediction for the
$\tau^-\bar{\nu}$ branching fraction of $(1.03\pm 0.29)\times 10^{-4}$. This
value is now consistent with the average. 

\begin{figure}[hbt]
\centering
\includegraphics[width=80mm]{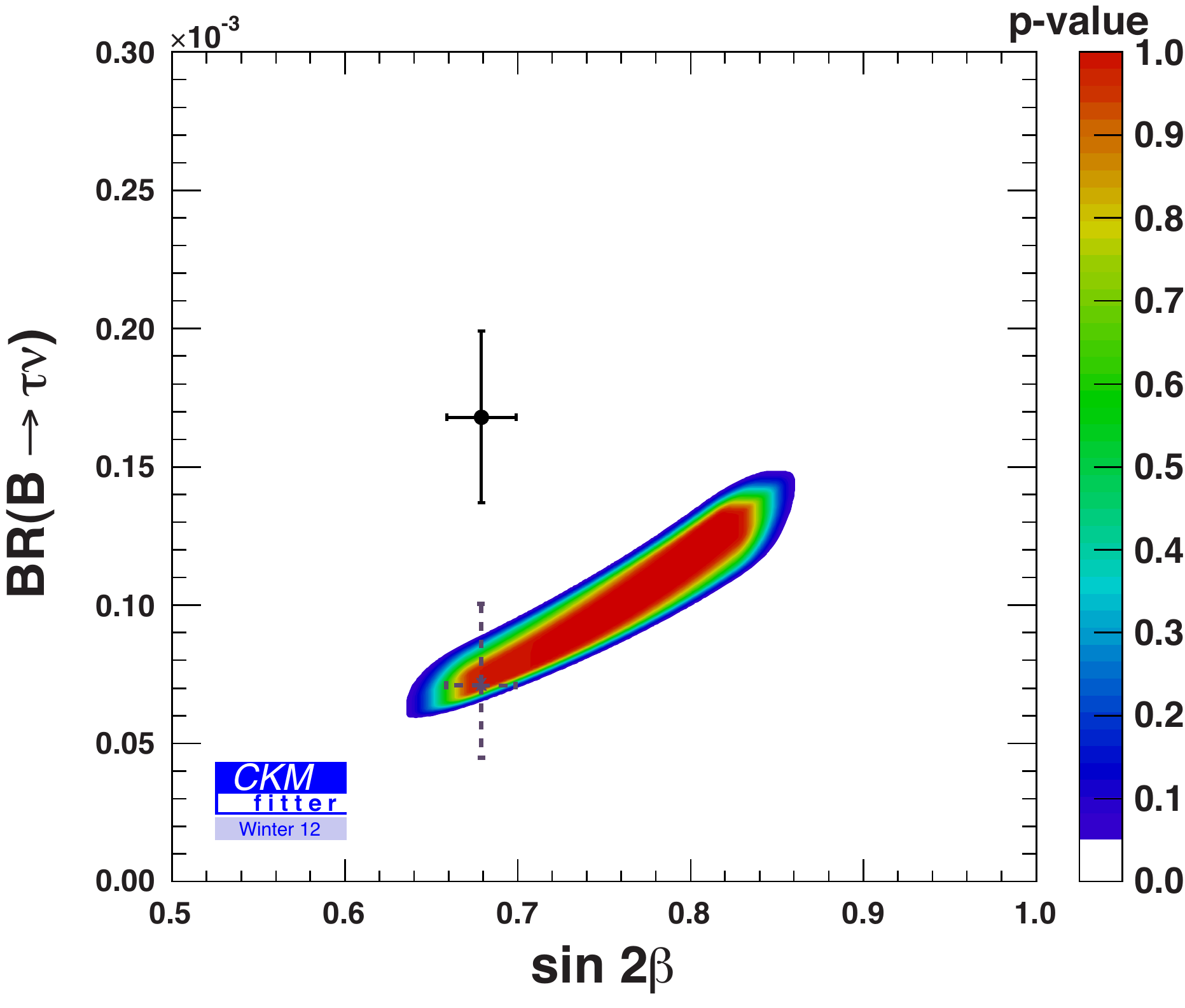}
\vspace{-4mm}
\caption{Measured versus predicted values of ${\cal{B}}(B^-\to\tau^-
\overline{\nu})$ versus $\sin2\beta$ from the CKM fitter group. The solid
(black) point with error bars shows the old (2012) measured average value, the
dashed (purple) point the new Belle measurement, while the predictions are in
colors, with the color being related to the confidence level. (Adopted from the
CKM Fitter group.)}
\label{sin2b_Btaunu}
\end{figure}

It is instructive to examine the correlation between the CKM angle $\beta$ and
${\cal{B}}(B^-\to\tau^-\bar{\nu})$. The CKM fitter group provides a fit to a
large number of measurements involving heavy quark transitions
\cite{CKMfitter}.  The black point in Fig.~\ref{sin2b_Btaunu} shows the
directly measured values from 2012, while the predictions from their fit
without the direct measurements are also shown.  There is about a factor of two
discrepancy between the old measured average value of ${\cal{B}}(B^-\to\tau^-
\overline{\nu})$ and the fit prediction. The (purple) dashed point shows the
new Belle measurement only, and is consistent with the prediction, as is the
new average.

\section{Charged pions and kaons}
We now discuss the determination of charged pion and kaon decay constants.  The
sum of branching fractions for $\pi^- \to \mu^- \bar \nu$ and $\pi^- \to \mu^-
\bar \nu \gamma$ is 99.98770(4)\%.  The two modes are difficult to separate
experimentally, so we use this sum, with Eq.~(\ref{equ_rate}) modified to
include photon emission and radiative corrections \cite{Marciano-Sirlin}.  The
branching fraction together with the lifetime 26.033(5) ns gives
$$
 f_{\pi^-} = (130.41\pm 0.03\pm 0.20)~{\rm MeV}~.
$$
\noindent The first uncertainty is due to the error on $|V_{ud}|$, 0.97425(22)
\cite{Vud}; the second is due to the higher-order corrections, and is much
larger.

Similarly, the sum of branching fractions for $K^- \to \mu^- \bar
\nu$ and $K^- \to \mu^- \bar \nu \gamma$ is 63.55(11)\%, and the
lifetime is 12.3840(193) ns \cite{Flavi}.  Measurements of semileptonic kaon
decays provide a value for the product $f_+(0)|V_{us}|$, where $f_+(0)$ is the
form-factor at zero four-momentum transfer between the initial state kaon and
the final state pion. We use a value for $f_+(0)|V_{us}|$ of 0.2163(5)
\cite{Flavi}. The $f_+(0)$ must be determined theoretically.
The two most recent determinations from lattice QCD are 0.9667(23)(33)
\cite{Baz13} and $0.9599(34)(^{+31}_{-43})$ \cite{Boyle10}, whose average
is $f_+(0) = 0.9638(30)$.
This is more precise than the classic Leutwyler-Roos calculation $f_+(0)=0.961
\pm 0.008$ \cite{LR}. The result is $|V_{us}|=0.2244(9)$, which is
consistent with the hyperon decay value of $0.2250(27)$ \cite{hyperon}.

Experimental branching ratios provide the ratio \cite{Dowdall}
$$
\frac{|V_{us}| f_{K^+}}{|V_{ud}|f_{\pi^-}} = 0.27598(35)(25)~,
$$
where the first error is due to branching fractions and the second is
due to electromagnetic corrections.  With $V_{ud} = 0.97425(22)$,
$f_{\pi^-}$ as given above, and $|V_{us}|=0.2244(9)$, we then find
$$
f_{K^-} =(156.2 \pm 0.2\pm 0.6\pm 0.3)~{\rm MeV}~.
$$

\noindent The first uncertainty is due to the error on $\Gamma$; the second is
due to the CKM factor $|V_{us}|$, and the third is due to the higher-order
corrections. The largest source of error in these corrections
depends on the QCD part, which is based on one calculation in the
large $N_c$ framework.  A large part of the additional uncertainty
vanishes in the ratio of the $K^-$ and $\pi^-$ decay constants, which is
$$
f_{K^-}/f_{\pi^-} = 1.198 \pm 0.002 \pm 0.005 \pm 0.001~.
$$

\noindent
The first uncertainty is due to the measured decay rates; the second is due to
the uncertainties on the CKM factors; the third is due to the errors in the
radiative correction ratio.
These measurements can be used in conjunction with calculations of
$f_K/f_{\pi}$ in order to find a value for $|V_{us}|/|V_{ud}|$ \cite{BillM}.  Recent
lattice predictions of $f_K/f_{\pi}$ are shown in Table~\ref{tab:fkfpi}. 

\begin{table}[htb]
\caption{Lattice calculations of $f_K/f_{\pi}$ and extracted values of $|V_{us}|/|V_{ud}|$.
\label{tab:fkfpi}}
\begin{center}
\begin{tabular}{ccc} \hline\hline
Group & $f_K/f_{\pi}$ & $|V_{us}|/|V_{ud}|$\\\hline
HPQCD \cite{Dowdall} & $1.1916 \pm 0.0021$ & 0.23160(54) \\
Laiho and Van de Water \cite{LV} & $1.202\pm 0.011\pm 0.013$& $-$\\
BMW \cite{fkfpi}& $1.192 \pm 0.007 \pm 0.006$&0.2315(19) \\
MILC \cite{Bazavov}& $1.1947 \pm0.0026\pm0.0037$& 0.2309(10)  \\
RBC/UKQCD \cite{RBC} & $1.204\pm 0.007\pm 0.025$&$-$\\
\hline\hline
\end{tabular}
\end{center}
\end{table}

These calculations are in agreement with our experimental average.  Together
with the precisely measured $|V_{ud}|$, these results can be used to find 
an independent measure of $|V_{us}|$ \cite{Jutt,Flavi}.

We gratefully acknowledge support of the U. S. National Science Foundation
and the U. S. Department of Energy through Grant No.\ DE-FG02-90ER40560.
We thank A. Khodjamirian, J. Laiho, W. Marciano, S. Narison, and Z.-G. Wang
for useful discussions and references.

\end{document}